\documentclass[11pt,twocolumn]{article}

\usepackage{arxiv}

\usepackage{amsmath,amssymb}

\usepackage{graphicx}
\usepackage{booktabs}
\usepackage{multirow}
\usepackage{tabularx}
\usepackage{colortbl}

\usepackage{float}
\usepackage{subcaption}

\usepackage{listings}
\usepackage[colorlinks=true,linkcolor=arxivblue,citecolor=arxivblue,urlcolor=arxivblue]{hyperref}
\usepackage{xurl}  

\usepackage[numbers,sort&compress]{natbib}

\newcommand{\realvuln}{\textsc{RealVuln}}

\title{\realvuln{}: Benchmarking Rule-Based,\\General-Purpose LLM, and Security-Specialized\\Scanners on Real-World Code}
\author{
  John Pellew \quad Faizan Raza \\[4pt]
  \texttt{john@kolega.ai} \quad \texttt{faizan@kolega.ai} \\
  \href{https://kolega.dev}{Kolega.Dev}
}
\date{March 31, 2026}

\begin{document}

\twocolumn[
  \begin{@twocolumnfalse}
    \maketitle
    \begin{abstract}
    \noindent
    How do security scanners perform on real-world code? We present \textsc{RealVuln}, the first open-source benchmark comparing Rule-Based SAST, General-Purpose LLMs, and Security-Specialized scanners on 26 intentionally vulnerable Python repositories (educational and Capture-The-Flag applications) with 796 hand-labeled entries (676 vulnerabilities, 120 false-positive traps). We test 15 scanners (3~Rule-Based SAST, 10~General-Purpose LLM, 2~Security-Specialized) and rank them by F3 score ($\beta{=}3$, weighting recall 9$\times$ over precision). A clear three-tier ranking emerges under all metrics. Under F3, the Security-Specialized scanner Kolega.Dev (73.0) leads, followed by the best General-Purpose LLM, Claude Sonnet~4.6 (51.7), which in turn scores nearly 3$\times$ higher than the best Rule-Based tool, Semgrep (17.7). Under F1, Sonnet~4.6 leads (60.9) with Kolega.Dev at 52.4. Rankings within tiers shift with $\beta$, but the three-tier hierarchy holds across all weightings. All code, ground-truth data, scanner outputs, and scoring scripts are released under an open-source license. An interactive dashboard is at \url{https://realvuln.kolega.dev/}. \textsc{RealVuln} is a living benchmark: versioned, community-driven, with a roadmap toward multi-language coverage.

    \end{abstract}
    \vspace{1em}
  \end{@twocolumnfalse}
]

\section{Introduction}

Static Application Security Testing (SAST) tools are a core part of secure
software development, but noise undermines their practical value.
The Ghost Security CAST report found a 99.5\% false positive rate for
command-injection findings in Python/Flask applications, with 91\% of all
2{,}166 findings across categories classified as noise~\cite{ghost2025cast}.
The pattern holds at scale: NIST's SATE~V and SATE~VI evaluations
found that only 8--30\% of tool warnings were security-relevant, depending on language and
tool~\cite{delaitre2018satev,delaitre2023satevi}. On the detection side,
the Fluid Attacks benchmark showed that the best fully automated tool found
only 22.7\% of known vulnerabilities, with an average F2 score (which weights recall over precision, where 100\% is perfect) of just 1.9\%
across 36 commercial and open-source scanners~\cite{fluidattacks2025}.
Practitioners, faced with hundreds of alerts they must manually triage,
increasingly ignore scanner output.

A new generation of LLM-powered security scanners promises to change this.
By using large language models for semantic code understanding rather than
pattern matching, these tools claim deeper reasoning about data flow, context,
and exploitability. But no open benchmark exists to test whether LLM-based
scanners actually deliver on these promises, or just trade one
set of failure modes for another. Existing benchmarks fall short on one or more
critical dimensions:

\begin{itemize}
  \item \textbf{Synthetic code.} The OWASP Benchmark~\cite{owaspbenchmark}
    consists of synthetic Java test cases that do not reflect the structure,
    complexity, or idioms of real-world applications.
  \item \textbf{Closed data.} SastBench~\cite{feiglin2026sastbench} is the
    closest prior work, but its dataset has not been publicly released
    and its evaluation focuses on triage agents rather than scanner accuracy.
  \item \textbf{Vendor self-assessment.} Benchmarks published by
    ZeroPath~\cite{zeropath2024}, Cycode~\cite{cycode2024benchmark}, and
    DryRun~\cite{dryrun2025sast} evaluate their own tools under vendor-controlled conditions. While some publish partial artifacts (ZeroPath releases benchmark code, DryRun uses public vulnerable apps), none provides a unified, reusable framework with open ground truth, scoring code, and multi-category comparison.
  \item \textbf{No false positive measurement.} Most benchmarks report
    detection rates alone, ignoring the false positive burden that dominates
    practitioner experience.
\end{itemize}

\noindent
We expand on these limitations in Section~\ref{sec:related-work}. The gap is
clear: no prior work systematically compares Rule-Based SAST, General-Purpose
LLMs, and Security-Specialized scanners on identical real-world code, measuring
not just what they find but also what they falsely flag.

This paper makes three contributions:

\begin{enumerate}
  \item \textbf{A systematic comparison of 15 scanners across three categories}
    on identical ground truth: 3 Rule-Based SAST tools (pattern matching),
    10 General-Purpose LLM scanner configurations spanning 7 model families
    (broad reasoning applied to security), and 2 Security-Specialized systems
    (purpose-built vulnerability detection).
  \item \textbf{The \textsc{RealVuln} benchmark} that enables this comparison: the
    first fully open-source SAST evaluation framework built from real-world code,
    comprising 26 intentionally vulnerable Python repositories (educational and Capture-The-Flag (CTF)
    applications such as PyGoat, DVPWA, and VAmPI) with 796
    hand-labeled findings and 120 false-positive traps. Version~1.0 covers
    Type~1 targets (intentionally vulnerable apps with high vulnerability density);
    Types~2--4 (production CVEs, vulnerable libraries, benchmark roll-ups) are
    planned for v2. We use F3 ($\beta{=}3$, recall weighted 9$\times$ over precision)
    as the primary metric but report F1 and F2 throughout; under F1, the
    top-ranked scanner changes (Section~\ref{sec:results}).
  \item \textbf{A living benchmark} with versioned scoring, a public
    dashboard, and a clear contribution path so that the community can add new
    scanners, repositories, and ground-truth labels over time.
\end{enumerate}

\paragraph{Disclosure.} One of the evaluated scanners (Kolega.Dev, \url{https://kolega.dev}) is developed by the authors of this paper. To mitigate bias, we open-source all ground-truth labels, scoring scripts, raw scanner outputs, and the evaluation harness, enabling full independent reproduction. We actively encourage third-party audits of our methodology and results.

We release all code, ground-truth data, scanner results, and scoring tooling at
\url{https://github.com/kolega-ai/Real-Vuln-Benchmark}, with a live interactive dashboard at \url{https://realvuln.kolega.dev/}.

\section{Background \& Related Work}
\label{sec:related-work}

Table~\ref{tab:benchmark-comparison} summarizes how RealVuln relates to prior
benchmarks across seven desirable properties. We discuss each category of prior
work below, highlighting the specific gaps that motivated our design.

\begin{table*}[t]
\caption{Comparison of SAST benchmarks. \emph{Real Code}: uses unmodified
  application source, not synthetic snippets. \emph{Open Data}: ground truth
  and scanner results publicly available. \emph{Open Scoring}: scoring code
  released for reproduction. \emph{FP Testing}: explicitly measures false
  positives. \emph{Multi-Scanner}: evaluates $\geq$3 scanners head-to-head.
  \emph{LLM Scanners}: includes LLM-based tools. \emph{Multi-Lang}: covers
  more than one programming language.}
\label{tab:benchmark-comparison}
\centering
\footnotesize
\begin{tabular}{@{}lccccccc@{}}
\toprule
\textbf{Benchmark} & \textbf{Real} & \textbf{Open} & \textbf{Open} & \textbf{FP} & \textbf{Multi-} & \textbf{LLM} & \textbf{Multi-} \\
 & \textbf{Code} & \textbf{Data} & \textbf{Scoring} & \textbf{Test} & \textbf{Scanner} & \textbf{Scan.} & \textbf{Lang} \\
\midrule
OWASP Benchmark~\cite{owaspbenchmark}
  & $\times$ & \checkmark & \checkmark & \checkmark & \checkmark & $\times$ & $\times$ \\
Juliet Test Suite~\cite{boland2012juliet}
  & $\times$ & \checkmark & $\times$ & $\times$ & $\times$ & $\times$ & $\times$ \\
SastBench~\cite{feiglin2026sastbench}
  & \checkmark & $\times$ & $\times$ & \checkmark & $\times$ & $\times$ & \checkmark \\
ZeroPath/XBOW~\cite{zeropath2024}
  & $\times$ & $\sim$ & $\times$ & \checkmark & \checkmark & \checkmark & $\times$ \\
Fluid Attacks~\cite{fluidattacks2025}
  & \checkmark & $\times$ & $\times$ & \checkmark & \checkmark & $\times$ & $\times$ \\
CASTLE~\cite{dubniczky2025castle}
  & $\times$ & \checkmark & \checkmark & $\times$ & \checkmark & \checkmark & \checkmark \\
Cycode~\cite{cycode2024benchmark}
  & \checkmark & $\times$ & $\times$ & $\times$ & \checkmark & $\times$ & \checkmark \\
DryRun~\cite{dryrun2025sast}
  & \checkmark & $\sim$ & $\times$ & $\times$ & \checkmark & $\times$ & \checkmark \\
\midrule
\textbf{RealVuln (ours)}
  & \checkmark & \checkmark & \checkmark & \checkmark & \checkmark & \checkmark & $\times$ \\
\bottomrule
\end{tabular}
\end{table*}

\paragraph{Synthetic benchmarks.}
The OWASP Benchmark~\cite{owaspbenchmark} (${\sim}$2{,}740 synthetic Java servlets) and the NIST Juliet Test Suite~\cite{boland2012juliet} (single-CWE C/C++/Java programs) are the most cited SAST evaluation suites. Both use short, self-contained test cases that let pattern-matching tools score well without understanding cross-file data flow or framework semantics, limiting their relevance to whole-application analysis.

\paragraph{Unreleased datasets.}
SastBench~\cite{feiglin2026sastbench} assembles 2{,}737 ground-truth samples across 38 languages from real CVEs, scored with the Matthews Correlation Coefficient. Its dataset is not publicly released, and it evaluates agentic \emph{triage} of SAST alerts rather than raw scanner detection accuracy. SastBench is complementary to RealVuln. It measures post-scan triage that we deliberately exclude, but cannot serve as an open benchmark for detection performance.

\paragraph{Vendor benchmarks.}
ZeroPath~\cite{zeropath2024} published benchmark code and SARIF results as a public GitHub fork of the XBOW validation benchmarks, testing against major SAST vendors, a genuine step toward transparency, though without reusable scoring infrastructure or ground-truth labels. Fluid Attacks~\cite{fluidattacks2025} tested 36 scanners against a large JS/TS application (the best achieved 22.7\% recall), but neither the source nor the ground truth is public. Cycode~\cite{cycode2024benchmark} ran Bearer CLI against open-source projects and published raw classification data, but the methodology is not fully documented for reproduction. DryRun~\cite{dryrun2025sast} tested against public intentionally-vulnerable applications with per-vulnerability breakdowns, providing partial reproducibility. The shared limitation across all vendor benchmarks is a structural conflict of interest: each vendor selects the test conditions, interprets results, and lacks a unified framework for independent re-scoring.

\paragraph{Dataset quality and label noise.}
Automated vulnerability datasets suffer from widespread label noise. Ding et al.~\cite{ding2024primevul} showed that a 7B model achieves 68\% F1 on BigVul but only 3\% on their cleaned PrimeVul dataset; CleanVul~\cite{li2025cleanvul} found over 40\% of entries in existing datasets mislabeled. CVEFixes~\cite{bhandari2021cvefixes} and ReposVul~\cite{wang2024reposvul} inherit the same tangled-commit problems. RealVuln takes the opposite position: 796 hand-labeled findings across 26 repositories, accepting lower scale in exchange for label quality that can be audited line by line.

\paragraph{Agentic and LLM-oriented benchmarks.}
SWE-Bench~\cite{jimenez2024swebench} introduced the approach of evaluating AI agents on real GitHub issues with reproducible harnesses. SWE-Bench Verified~\cite{openai2025swebenchverified} further demonstrated the value of human-validated labels. CVE-Bench~\cite{zhu2025cvebench} measures whether agents can exploit real CVEs in containerized apps, highlighting knowledge-cutoff confounds. CASTLE~\cite{dubniczky2025castle} evaluates 25 tools (13 static analyzers, 10 LLMs, and 2 formal verification tools) on CWE detection but uses synthetic micro-benchmarks. Yildiz et al.~\cite{yildiz2025jitvul} found that agentic approaches outperformed function-level prompting for vulnerability detection but noted that agents still exhibit inconsistencies requiring further refinement.

\paragraph{Evaluation metrics in prior work.}
Prior benchmarks use a range of metrics, none of which adequately captures the risk profile of security scanning. The OWASP Benchmark reports True Positive Rate (TPR) and False Positive Rate (FPR) independently, leaving practitioners to weigh the tradeoff themselves. Youden's J statistic ($J = \text{TPR} - \text{FPR}$) combines both into a single number but weights them equally, treating a missed vulnerability and a false alarm as equivalent costs. In practice, they are not: a single missed vulnerability can lead to a breach, while a false positive costs analyst time. SastBench~\cite{feiglin2026sastbench} uses the Matthews Correlation Coefficient (MCC), which is well-suited for balanced classification but does not let the evaluator express a preference for recall over precision. Fluid Attacks~\cite{fluidattacks2025} reports F2, the first recall-weighted metric in this space, reflecting the intuition that missed findings matter more than false alarms. CASTLE~\cite{dubniczky2025castle} proposes its own composite CASTLE Score. PrimeVul~\cite{ding2024primevul} and most academic vulnerability detection papers default to F1, which weights precision and recall equally.

F1 is the most common metric in the literature, but it does not reflect the asymmetry of real-world security risk. An attacker needs one undetected flaw to compromise a system. A defender needs to find every flaw. Weighting precision and recall equally implicitly assumes the cost of a missed vulnerability equals the cost of a false positive. For development-time scanning where noisy alerts erode trust, that assumption may hold. For production security, compliance audits, and regulated industries, it does not. Industry experience shows that developers stop trusting tools when false positive rates exceed roughly 30\%, but that threshold is a usability concern, not a security one. A tool with 30\% FPR and 90\% recall protects more code than one with 5\% FPR and 25\% recall.

F2 ($\beta{=}2$, recall weighted $4\times$) moves in the right direction and has precedent in Fluid Attacks. We argue that F3 ($\beta{=}3$, recall weighted $9\times$) better reflects the cost asymmetry for production and high-risk environments. Section~\ref{sec:scoring} details our metric design and this rationale.

\paragraph{Positioning and limitations of RealVuln.}
No prior benchmark systematically compares all three scanner
categories (Rule-Based SAST, General-Purpose LLMs, and Security-Specialized
systems) on identical ground truth. As Table~\ref{tab:benchmark-comparison}
shows, RealVuln is the only benchmark that simultaneously uses real code,
releases all data and scoring code, tests for false positives, and includes
scanners from all three categories.
We are forthright about its limitations in v1. First, RealVuln covers
Python only. This enables depth (Django, Flask, FastAPI, and raw Python
idioms are all represented), it cannot speak to scanner performance on memory-unsafe
languages or compiled ecosystems. Second, our ground truth covers only
OWASP-style application vulnerabilities (injection, XSS, authentication
flaws), but we do not yet include supply-chain, configuration, or
infrastructure-as-code findings. Third, we are ourselves a vendor (Kolega.Dev),
which creates the same conflict of interest we criticize above. We mitigate
this by open-sourcing \emph{everything}: ground truth labels, all scanner
result files, the scoring pipeline, and the interactive dashboard, so that any
researcher can audit, challenge, or extend our results. We view transparency
not as a defense against bias but as an invitation for the community to find it.

\section{Benchmark Design}
\label{sec:benchmark-design}

This section describes the design of \textsc{RealVuln}~v1.0: the taxonomy of target types it recognizes, how the dataset was constructed and labeled, the algorithm used to match scanner findings against ground truth, and the scoring metrics reported.

\subsection{Target Type Taxonomy}
\label{sec:target-types}

We define five \emph{target types} to classify the code under test.
Version~1.0 of \textsc{RealVuln} covers only Type~1.

\medskip
\noindent\textbf{Type 1: Intentionally vulnerable applications.}
Open-source projects built \emph{on purpose} with security flaws for educational or CTF use (e.g., PyGoat, DVPWA, VAmPI).
These projects offer high vulnerability density and diverse CWE coverage, making them ideal for initial calibration.

\medskip
\noindent Types 2--5 (production CVEs, vulnerable libraries, benchmark roll-ups, and academic reproductions) are planned for future versions and defined precisely in the roadmap (Section~\ref{sec:living-benchmark}).

\subsection{Dataset Construction}
\label{sec:dataset-construction}

\subsubsection{Repository Selection.}
We collected 26 intentionally vulnerable Python repositories from GitHub.
Selection criteria were: (1)~the project must be a multi-file application with a meaningful code structure (single-file toy scripts were excluded), (2)~the project must use a recognizable web framework, and (3)~the project should contribute CWE families not already saturated in the dataset.
Each repository is pinned to a specific commit SHA recorded in \texttt{benchmark-manifest.json}, ensuring that all scanners analyze identical source trees.

The 26~repositories span five web frameworks: Flask (15~repos), Django (3), FastAPI (3), aiohttp (1), and Tornado (1), with 3~repos using lightweight or custom HTTP stacks.
Table~\ref{tab:dataset-stats} summarizes the dataset.

\begin{table}[t]
  \centering
  \caption{Dataset statistics for \textsc{RealVuln}~v1.0.}
  \label{tab:dataset-stats}
  \footnotesize
  \begin{tabular}{@{}lr@{}}
    \toprule
    \textbf{Statistic} & \textbf{Value} \\
    \midrule
    Repositories          & 26  \\
    Total findings        & 796 \\
    \quad Vulnerabilities & 676 \\
    \quad FP traps        & 120 \\
    Frameworks            & 5   \\
    CWE families          & 18  \\
    Language              & Python \\
    Target type           & Type~1 \\
    \bottomrule
  \end{tabular}
\end{table}

\subsubsection{Ground Truth Labeling.}
Every finding in the ground truth was produced by manual review.
Each labeled entry records the following fields:

\begin{itemize}
  \item \textbf{Identity:} a unique \texttt{id} and a boolean \texttt{is\_vulnerable} flag.
  \item \textbf{CWE classification:} a \texttt{primary\_cwe} reflecting the most precise weakness, and an \texttt{acceptable\_cwes} list enumerating alternative CWE identifiers that a scanner may reasonably report for the same flaw (e.g., both CWE-89 and CWE-943 for a SQL injection that uses an ORM raw query).
  \item \textbf{Location:} the file path and a \texttt{start\_line}/\texttt{end\_line} range pinpointing the vulnerable code.
  \item \textbf{Severity:} one of \texttt{critical}, \texttt{high}, \texttt{medium}, or \texttt{low}.
  \item \textbf{Evidence:} the source of the annotation (\texttt{manual\_review}, \texttt{cve\_id}, or \texttt{walkthrough}) and a free-text description explaining \emph{why} the code is or is not vulnerable.
\end{itemize}

\noindent
The full ground-truth schema and example entries are available in the repository.\footnote{\url{https://github.com/kolega-ai/Real-Vuln-Benchmark/blob/main/ground-truth/realvuln-djangoat/ground-truth.json}}

\subsubsection{Authorship Metadata.}
All 26~repositories are tagged \texttt{human\_authored} with high confidence.
These projects predate the widespread availability of large language models for code generation, which is relevant for mitigating data contamination concerns when evaluating LLM-based scanners: the ground truth code was not plausibly generated by the models under test.

\subsubsection{False Positive Traps.}
Of the 796~labeled findings, 120 (15.1\%) are \emph{FP~traps}: code patterns that appear suspicious but are demonstrably safe upon manual analysis.
For example, a login function that passes user input to SQLAlchemy's \texttt{filter\_by()} method superficially resembles SQL injection, but the ORM automatically parameterizes the query.
If a scanner flags such an entry, the match is classified as a false positive and penalized in scoring.
FP~traps serve two purposes: (1)~they test a scanner's ability to distinguish truly vulnerable code from safe-but-suspicious patterns, and (2)~they provide true negatives for computing specificity-related metrics such as the false positive rate.

\subsection{Matching Algorithm}
\label{sec:matching}

Each scanner finding is matched against the ground truth on three fields: (1)~\textbf{file path} (exact match after normalization), (2)~\textbf{CWE} (the scanner's CWE must appear in the ground truth entry's \texttt{acceptable\_cwes} list), and (3)~\textbf{line number} (within $\pm$10 lines of the ground truth range). When a finding matches multiple entries, we prefer real vulnerabilities over FP traps, giving the scanner the benefit of the doubt. Each ground truth entry is consumed at most once. Unmatched findings are false positives, unmatched vulnerable entries are false negatives, and unmatched traps are true negatives.

\paragraph{Alternative locations.}
Some scanners report attack chains rather than single-point locations: a single audit result may reference multiple files (the vulnerable handler, its route definition, model layers, etc.) while describing one or more vulnerabilities. To handle this fairly, a finding may declare \emph{alternative locations}: additional (file,~line) pairs that the matcher tries if the primary location does not match. If any location matches, the finding is not a false positive. When a single finding's alternatives match multiple distinct ground truth entries, each match is credited as a separate true positive, since the scanner genuinely identified multiple vulnerabilities in one report. This prevents penalizing chain-of-evidence reporting styles while still counting one false positive per unmatched finding.

The full matching implementation is available in the repository.\footnote{\url{https://github.com/kolega-ai/Real-Vuln-Benchmark/blob/main/scorer/matcher.py}}

\subsection{Scoring}
\label{sec:scoring}

\subsubsection{Primary Metric: $F_3$ Score.}

Two base metrics underpin all scoring:

\begin{itemize}
  \item \textbf{Precision:} of everything the scanner flagged, what fraction was actually vulnerable? High precision means less noise for analysts to triage.
  \item \textbf{Recall:} of all real vulnerabilities, what fraction did the scanner find? High recall means fewer missed vulnerabilities.
\end{itemize}

\noindent
How these two metrics are combined determines what a benchmark rewards. The $F_\beta$ family generalizes the harmonic mean of precision and recall with a parameter $\beta$ that controls the tradeoff: $\beta = 1$ weights them equally, higher $\beta$ favors recall. Section~\ref{sec:related-work} reviews the metrics used in prior work and argues that equal weighting (F1, Youden's J) does not reflect the cost asymmetry in security, where missed vulnerabilities carry far greater risk than false positives.

Based on that argument, we recommend $F_3$ ($\beta{=}3$) as the primary metric, scaled to $[0, 100]$:
\begin{equation}
  F_3 = \frac{10 \cdot P \cdot R}{9P + R} \times 100
  \label{eq:f3}
\end{equation}
Setting $\beta = 3$ weights recall nine times more heavily than precision. A scanner that finds 90\% of vulnerabilities with 30\% precision scores higher than one that finds 40\% with 90\% precision. We report F1 and F2 alongside F3 throughout the paper so readers can evaluate rankings under their own preferred weighting. The full leaderboard data in the repository allows re-ranking under any $F_\beta$.

\subsubsection{Secondary Metric: $F_2$ Score.}
We additionally report the $F_2$ score:
\begin{equation}
  F_2 = \frac{5 \cdot P \cdot R}{4P + R} \times 100
  \label{eq:f2}
\end{equation}
Setting $\beta = 2$ weights recall four times more heavily than precision. $F_2$ remains appropriate for lower-risk projects or development-time scanning where noise reduction matters more, but $F_3$ is the recommended default for production and high-risk environments.

\subsubsection{Per-CWE-Family and Per-Severity Breakdowns.}
Metrics are also computed per CWE family (18 families grouping related CWEs, e.g., SQL Injection covers CWE-89, CWE-564, CWE-943) and per severity level (critical, high, medium, low), revealing category-level blind spots.

\subsubsection{Scoring Modes and Multi-Run Aggregation.}
\textsc{RealVuln} supports two aggregation modes for multi-repo evaluation:

\begin{description}
  \item[\texttt{micro}] pools TP, FP, FN, and TN counts across all repositories before computing metrics. Repositories where the scanner produced no output (e.g., due to a timeout or parsing failure) are simply omitted from the pool.
  \item[\texttt{strict\_micro}] is identical except that repositories with no scanner output are treated as if the scanner reported zero findings: every vulnerable GT entry in that repo becomes a false negative. This mode penalizes incomplete runs and is the conservative primary reporting mode in this paper. All headline scores are F3 (strict).
\end{description}

\noindent
When a scanner is run multiple times on the same repository (e.g., with different prompts or configurations), each run is scored independently.
Per-repo scores are then aggregated across runs using micro-averaging over the pooled confusion matrix.

\subsubsection{Reproducibility.}
The file \texttt{benchmark-manifest.json} locks three pieces of information necessary for exact reproduction: (1)~the content hash of the ground truth directory, (2)~the commit SHA for each of the 26~repositories, and (3)~a hash of the default prompt version used for LLM-based scanners.
Any change to the ground truth labels, repository code, or prompt template produces a new manifest, making it straightforward to detect drift between benchmark versions.

\section{Experimental Setup}
\label{sec:experimental-setup}

We evaluate 15 scanner configurations spanning three categories:
\emph{Rule-Based SAST} tools, \emph{General-Purpose LLM} scanners, and
two \emph{Security-Specialized} systems. This taxonomy reflects a
fundamental difference in approach: Rule-Based tools match syntactic
patterns, General-Purpose LLMs apply broad reasoning capabilities to a
security task they were not specifically built for, and
Security-Specialized tools combine LLM reasoning with domain-specific
architecture purpose-built for vulnerability detection.
All scanners are run against identical pinned commits of the 26
repositories in the RealVuln corpus.

\subsection{Scanners Evaluated}
\label{sec:scanners}

\paragraph{Rule-Based SAST (3 tools).}
We evaluate three widely deployed static analyzers: \textbf{Semgrep}~\cite{semgrep} (open-source, Rule-Based), \textbf{Snyk Code}~\cite{snyk} (commercial, pattern + dataflow), and \textbf{SonarQube}~\cite{sonarqube} (community edition, pattern-based). All three were run with default configurations against every repository in the benchmark.

\paragraph{General-purpose LLM scanners (10 configurations, 10 distinct models).}
We evaluate models from Anthropic (Claude Haiku~4.5, Sonnet~4.6, Opus~4.6), Google (Gemini~3.1~Pro), xAI (Grok~3, Grok~4.20~Reasoning), and four additional providers (Kimi~K2.5, GLM-5, Minimax~M2.7, Qwen~3.5~397B). All models use agentic mode, receive the same shared prompt (Section~\ref{sec:llm-modes}), and produce structured JSON output.

\paragraph{Security-Specialized scanners (2).}
\textbf{Kolega.Dev v0.0.1} is, to our knowledge, the first purpose-built
security scanner that combines LLM reasoning with domain-specific
architecture designed specifically for vulnerability detection. Unlike
the General-Purpose LLMs above, which receive a security prompt but
were not built for this task, Kolega.Dev integrates security-specific
analysis pipelines at the architectural level. Its scan results are
published in the repository. Its internals are proprietary.

\textbf{GitHub SecLab Taskflow Agent v1}~\cite{seclab-taskflow} is an open-source, multi-stage agentic auditor developed by GitHub Security Lab. In our evaluation, we configured it to use GPT-5.4 via a LiteLLM proxy for all analysis stages. The agent's architecture is model-agnostic and supports other backends. It performs threat-model-driven security analysis: for each repository it identifies 3--5 threat categories, then conducts deep targeted audits of each. Unlike Kolega.Dev's broad-coverage architecture, the SecLab agent is designed for analytical depth rather than breadth, auditing a narrow set of threat categories per scan rather than enumerating all vulnerability classes.

\medskip
\noindent
Table~\ref{tab:scanners} summarizes all evaluated scanners.

\begin{table*}[t]
  \caption{Scanner classification. \emph{Rule} = Rule-Based SAST;
    \emph{GP-LLM} = General-Purpose LLM; \emph{Sec.-spec.} = Security-Specialized.}
  \label{tab:scanners}
  \centering
  \footnotesize
  \begin{tabular}{@{}llll@{}}
    \toprule
    \textbf{Scanner} & \textbf{Cat.} & \textbf{Mode} & \textbf{OSS} \\
    \midrule
    Semgrep          & Rule  & Rule-Based         & Yes \\
    Snyk             & Rule  & Pattern+dataflow   & No  \\
    SonarQube        & Rule  & Pattern-based      & Yes \\
    \midrule
    Claude Haiku 4.5    & GP-LLM & Agentic       & No \\
    Claude Sonnet 4.6   & GP-LLM & Agentic       & No \\
    Claude Opus 4.6     & GP-LLM & Agentic       & No \\
    Gemini 3.1 Pro      & GP-LLM & Agentic       & No \\
    Grok 3              & GP-LLM & Agentic       & No \\
    Grok 4.20 Reasoning & GP-LLM & Agentic       & No \\
    Kimi K2.5           & GP-LLM & Agentic       & No \\
    GLM-5               & GP-LLM & Agentic       & No \\
    Minimax M2.7        & GP-LLM & Agentic       & No \\
    Qwen 3.5 397B       & GP-LLM & Agentic       & No \\
    \midrule
    Kolega.Dev v0.0.1  & Sec.-spec. & Purpose-built & No \\
    SecLab Agent v1 & Sec.-spec. & Agentic (multi-stage) & Yes \\
    \bottomrule
  \end{tabular}
\end{table*}

\subsection{LLM Evaluation Mode}
\label{sec:llm-modes}

All LLM-based scanners are evaluated in \textbf{agentic} mode.
The model operates within a tool-use loop, with access to file reading,
\texttt{grep}-style search, and shell execution tools. It can
iteratively explore the codebase, form hypotheses about data flow, and
refine its findings across multiple turns. This mode closely mirrors
how a human auditor would approach the task.

\paragraph{Shared prompt.}
All LLM configurations receive the same prompt template, identified by
its content hash (\texttt{sha256:828b00245b42}). The prompt instructs
the model to act as a security auditor, specifies the expected JSON
output schema (CWE identifier, file path, line number, severity), and
provides minimal guidance on vulnerability categories. Per-model prompt
tuning is intentionally avoided: while a shared prompt may
advantage or disadvantage particular models, it ensures that differences
in performance reflect model capability rather than prompt engineering
effort. The full prompt is published in the benchmark repository.

\paragraph{Output schema.}
Each LLM must return a JSON array of findings, where each finding
contains: a CWE identifier (\texttt{cwe}), the affected file path
(\texttt{file}), a line number (\texttt{line}), and a severity label
(\texttt{severity}). When a model returns malformed JSON (e.g., trailing commas, unescaped strings), we attempt automated repair using GPT-5 before discarding, giving each scanner the benefit of the doubt. Findings that still do not conform after repair are discarded.

\paragraph{Cost tracking.}
We record input and output token counts for every LLM invocation and
compute per-scan costs using provider pricing as of March~2026. This
enables cost-efficiency analysis (e.g., F3 score per dollar) in
Section~\ref{sec:results}, which is critical for practitioners
evaluating deployment trade-offs.

\subsection{Reproducibility}
\label{sec:reproducibility}

All artefacts needed to reproduce the evaluation are committed to the public repository: raw scanner output for every (repository, scanner) pair, a version-locked \texttt{benchmark-manifest.json} recording ground-truth hashes and commit SHAs, and the scoring pipeline itself. Any user can re-derive all metrics via \texttt{python score.py --repo <repo> --all-scanners} and regenerate the interactive dashboard via \texttt{python dashboard.py --scanner-group all}.

\section{Results}
\label{sec:results}

We now present the comparative evaluation of all 15 scanners, evaluated on 26 intentionally vulnerable Python repositories (${\sim}$20{,}000 lines of code total) containing 796 ground-truth entries (676 true vulnerabilities and 120 false-positive traps). All 26 repositories are Type~1 targets: educational and Capture-The-Flag (CTF) applications (e.g., PyGoat, DVPWA, VAmPI) designed with high vulnerability density. This means an average of roughly 26 vulnerabilities per repository, which is expected for this category but higher than typical production code. Ground-truth labeling applies production-grade standards even to teaching code: if an API lacks authentication by design, it is still flagged, because the benchmark evaluates what production scanners should catch. These are deliberately small repositories, representative of microservices and web applications rather than large monoliths, which makes them tractable for both LLM context windows and manual ground-truth labeling. Types~2--4 (production CVEs, vulnerable libraries, benchmark roll-ups) are planned for v2 and will exhibit realistic vulnerability densities. All results are also available as an interactive dashboard at \url{https://realvuln.kolega.dev/}.

All scores are F3 (strict) unless noted otherwise. Metric definitions appear in Section~\ref{sec:scoring}. We report F1 and F2 alongside F3 throughout so readers can evaluate rankings under different precision--recall tradeoffs.

\paragraph{Metric sensitivity.} The choice of $\beta$ changes the top-ranked scanner. Under F1 (equal weighting), Claude Sonnet~4.6 leads at 60.9, with Kolega.Dev at 52.4 (6th). Under F2, Kolega.Dev leads at 66.5, followed by Sonnet at 53.7. Under our primary metric F3 ($\beta{=}3$, recall weighted 9$\times$), Kolega.Dev leads at 73.0, followed by Sonnet at 51.7. The three-tier hierarchy (Security-Specialized $>$ GP-LLM $>$ Rule-Based SAST) holds under all three metrics, but within-tier rankings shift with $\beta$. We use F3 as the primary metric for reasons detailed in Section~\ref{sec:scoring}: in security, missed vulnerabilities carry asymmetric risk relative to false positives. Readers who weight precision more heavily should consult the F1 column.

\subsection{Aggregate Rankings}
\label{sec:results:aggregate}

Table~\ref{tab:leaderboard} presents the full leaderboard ranked by strict F3.

\begin{table*}[t]
\centering
\caption{RealVuln leaderboard ranked by strict F3 (primary metric). F1 and F2 are shown for comparison. Under F1, Claude Sonnet~4.6 leads (60.9); under recall-weighted F3, Kolega.Dev leads (73.0). \textbf{Category}: Rule = Rule-Based SAST; GP-LLM = General-Purpose LLM; Sec.-spec.\ = Security-Specialized. Metric definitions are in Section~\ref{sec:scoring}.}
\label{tab:leaderboard}
\small
\begin{tabular}{@{}llrrrrrc@{}}
\toprule
\textbf{Scanner} & \textbf{Cat.} & \textbf{F1} & \textbf{F2} & \textbf{F3} & \textbf{Prec} & \textbf{Recall} & \textbf{Repos} \\
\midrule
Kolega.Dev-v0.0.1                   & Sec.-spec. & 52.4 & \textbf{66.5} & \textbf{73.0} & 0.388 & \textbf{0.809} & 26/26 \\
seclab-taskflow-agent-v1            & Sec.-spec. & 13.7 &  9.3 & 8.4           & 0.605 & 0.077          & 26/26 \\
\midrule
claude-sonnet-4.6-agentic-v1        & GP-LLM    & \textbf{60.9} & 53.7 & 51.7          & 0.785 & 0.498          & 23/26 \\
gemini-3.1-pro-agentic-v1           & GP-LLM    & 60.1 & 53.0 & 51.0          & 0.774 & 0.491          & 24/26 \\
claude-opus-4.6-agentic-v1          & GP-LLM    & 57.8 & 49.8 & 47.7          & 0.790 & 0.456          & 19/26 \\
kimi-k2.5-agentic-v1                & GP-LLM    & 54.3 & 48.3 & 46.6          & 0.683 & 0.450          & 24/26 \\
glm-5-agentic-v1                    & GP-LLM    & 55.5 & 47.9 & 45.8          & 0.753 & 0.439          & 22/26 \\
minimax-m2.7-agentic-v1             & GP-LLM    & 48.7 & 41.0 & 39.0          & 0.707 & 0.371          & 22/26 \\
qwen-3.5-397b-agentic-v1            & GP-LLM    & 46.0 & 39.8 & 38.1          & 0.618 & 0.366          & 24/26 \\
claude-haiku-4.5-agentic-v1         & GP-LLM    & 47.9 & 39.4 & 37.2          & 0.748 & 0.352          & 24/26 \\
grok-4.20-reasoning-agentic-v1      & GP-LLM    & 41.0 & 30.7 & 28.4          & \textbf{0.927} & 0.263 & 24/26 \\
grok-3-agentic-v1                   & GP-LLM    & 31.9 & 23.3 & 21.3          & 0.837 & 0.197          & 21/26 \\
\midrule
semgrep                             & Rule SAST   & 18.9 & 18.0 & 17.7          & 0.205 & 0.175          & 25/26 \\
snyk                                & Rule SAST   & 21.0 & 18.2 & 17.4          & 0.282 & 0.167          & 25/26 \\
sonarqube                           & Rule SAST   & 11.8 &  7.9 & 7.1           & 0.611 & 0.065          & 26/26 \\
\bottomrule
\end{tabular}
\end{table*}

\paragraph{Headline findings.}
The results reveal a clear three-tier hierarchy.
\textbf{The Security-Specialized scanner (Kolega.Dev) leads the benchmark at F3\,=\,73.0 (strict)}, achieving the highest recall of any scanner (0.809). Its purpose-built architecture for vulnerability detection delivers broad coverage that neither Rule-Based nor General-Purpose LLM approaches achieve alone.
Its precision of 0.388 reflects the fundamental tradeoff at the recall-optimized end of the spectrum: Kolega.Dev flags more code for review, but in doing so catches over 80\% of true vulnerabilities. In production security, where attackers need only a single undetected flaw to compromise a system, this tradeoff is strongly favorable. The cost of reviewing false positives is measured in analyst minutes. The cost of a missed vulnerability can be measured in data breaches.

\textbf{General-purpose LLM scanners form a competitive middle tier}, with Claude Sonnet~4.6 achieving the best strict F3 (51.7), narrowly ahead of Gemini~3.1~Pro (51.0).
Both balance precision near 0.78 with recall near 0.49.
Notably, Claude Opus~4.6 drops to F3\,=\,47.7 on the strict metric because it completed only 19 of 26 repos, a \textbf{27\% repo failure rate} that suggests frontier reasoning models may struggle with the time constraints of real-world scanning workloads.

\textbf{Rule-Based SAST tools occupy the bottom tier}, with Semgrep (F3\,=\,17.7), Snyk (17.4), and SonarQube (7.1) all scoring below every General-Purpose LLM scanner.

The second Security-Specialized scanner, the open-source GitHub SecLab Taskflow Agent~\cite{seclab-taskflow}, scores F3\,=\,8.4, below all Rule-Based SAST tools. It uses GPT-5.4 and a multi-stage threat-modeling pipeline but finds fewer than 8\% of ground-truth vulnerabilities. Its precision is high (0.605) because its audit reports are thorough, but its recall is limited by design: the agent selects only 3--5 threat categories per repository, leaving entire vulnerability classes unexamined. ``Security-Specialized'' architecture alone is not enough. Broad detection coverage matters as much as analytical depth.

At the conservative extreme, Grok~4.20~Reasoning achieves the highest precision (0.927) but recalls only 26.3\% of vulnerabilities (F3\,=\,28.4).

\textbf{Rule-Based SAST tools show substantially lower detection rates.}
All three score below F3\,=\,18: Semgrep (17.7), Snyk (17.4), and SonarQube (7.1, detecting only 6.5\% of vulnerabilities despite reasonable precision).
Most agentic LLM scanners achieve \textbf{2--3$\times$ higher F3} than these established tools, suggesting that any organization relying solely on Rule-Based SAST has substantial undetected exposure.

\subsection{Precision vs.\ Recall}
\label{sec:results:prec-recall}

\begin{figure}[t]
\centering
\includegraphics[width=\columnwidth]{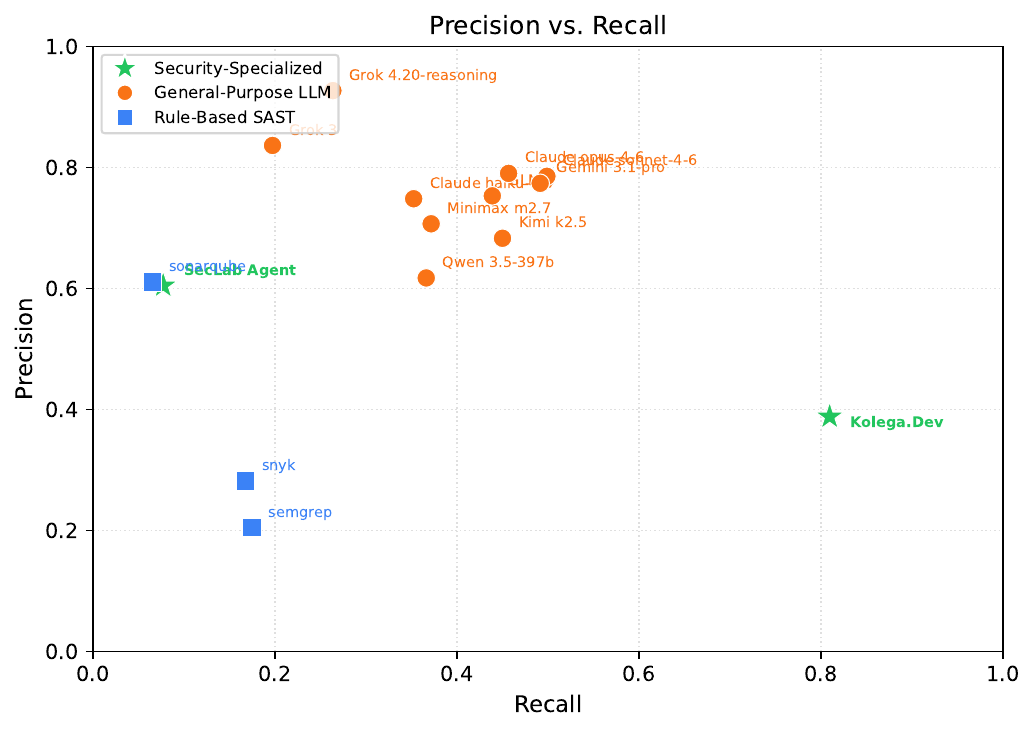}
\caption{Precision vs.\ recall for all 15 scanners, colored by category. Kolega.Dev occupies the high-recall right side (0.809) with moderate precision, GP-LLMs cluster in the center with higher precision but lower recall, and Rule-Based SAST tools are confined to the low-recall left. The two Security-Specialized scanners sit at opposite extremes: Kolega.Dev (high recall) and SecLab Agent (high precision, very low recall).}
\label{fig:prec-recall}
\end{figure}

Figure~\ref{fig:prec-recall} visualizes the precision--recall tradeoff across all scanners. Points toward the upper-right corner represent better performance, but the optimal tradeoff depends on context. Our F3 metric favors recall (upper-right and right), but teams with limited triage capacity may prefer higher-precision scanners (upper region). The full leaderboard data allows re-ranking under any $F_\beta$ weighting. The three-tier hierarchy is immediately apparent: Kolega.Dev occupies the high-recall region, GP-LLM scanners cluster in the center with higher precision but substantially lower recall, and Rule-Based SAST tools are confined to the bottom of the plot with both low precision and low recall.

\subsection{Cost-Efficiency Analysis}
\label{sec:results:cost}

Table~\ref{tab:cost-efficiency} compares cost and detection performance across all scanner categories. Costs are normalized to \$/100k lines of code to enable direct comparison across different pricing models.

\begin{table}[t]
\centering
\caption{Cost-efficiency comparison. Costs normalized to \$/100k LOC. Rule-Based SAST tools are free/open-source. Kolega.Dev pricing is \$25/100k LOC (published rate).}
\label{tab:cost-efficiency}
\footnotesize
\begin{tabular}{@{}llrr@{}}
\toprule
\textbf{Scanner} & \textbf{Cat.} & \textbf{F3} & \textbf{\$/100k LOC} \\
\midrule
Kolega.Dev v0.0.1    & Sec.-spec.  & \textbf{73.0} & \textbf{25} \\
SecLab Agent v1      & Sec.-spec.  & 8.4            & 125 \\
\midrule
Claude Sonnet 4.6    & GP-LLM      & 51.7 & 83 \\
Gemini 3.1 Pro       & GP-LLM      & 51.0 & 136 \\
Claude Opus 4.6      & GP-LLM      & 47.7 & 123 \\
Kimi K2.5            & GP-LLM      & 46.6 & 11 \\
GLM-5                & GP-LLM      & 45.8 & 34 \\
Qwen 3.5 397B        & GP-LLM      & 38.1 & 16 \\
Minimax M2.7         & GP-LLM      & 39.0 & 8 \\
Claude Haiku 4.5     & GP-LLM      & 37.2 & 26 \\
Grok 4.20 Reasoning  & GP-LLM      & 28.4 & 84 \\
Grok 3               & GP-LLM      & 21.3 & 28 \\
\midrule
Semgrep              & Rule SAST   & 17.7 & Variable \\
Snyk                 & Rule SAST   & 17.4 & Variable \\
SonarQube            & Rule SAST   & 7.1  & Variable \\
\bottomrule
\end{tabular}
\end{table}

\textbf{Kolega.Dev achieves the best cost-efficiency tradeoff.} At \$25 per 100k LOC, it achieves the highest F3 (73.0) at a fraction of the cost of General-Purpose LLM scanners. The best-performing GP-LLM (Claude Sonnet~4.6, F3\,=\,51.7) costs \$83/100k LOC, 3.3$\times$ more expensive for 29\% lower F3. The most expensive scanner (Gemini~3.1~Pro, \$136/100k LOC) achieves F3\,=\,51.0, costing 5.4$\times$ more than Kolega.Dev for 30\% lower detection.

Among GP-LLM scanners, Kimi~K2.5 (\$11/100k LOC, F3\,=\,46.6) and Minimax~M2.7 (\$8/100k LOC, F3\,=\,39.0) offer the best value, but even these budget options score 36--47\% below Kolega.Dev.

Rule-Based SAST tools have variable pricing but, as shown in Section~\ref{sec:results:aggregate}, detect fewer than 18\% of vulnerabilities.

\subsection{F3 Score vs.\ Cost}
\label{sec:results:f3-cost}

\begin{figure}[t]
\centering
\includegraphics[width=\columnwidth]{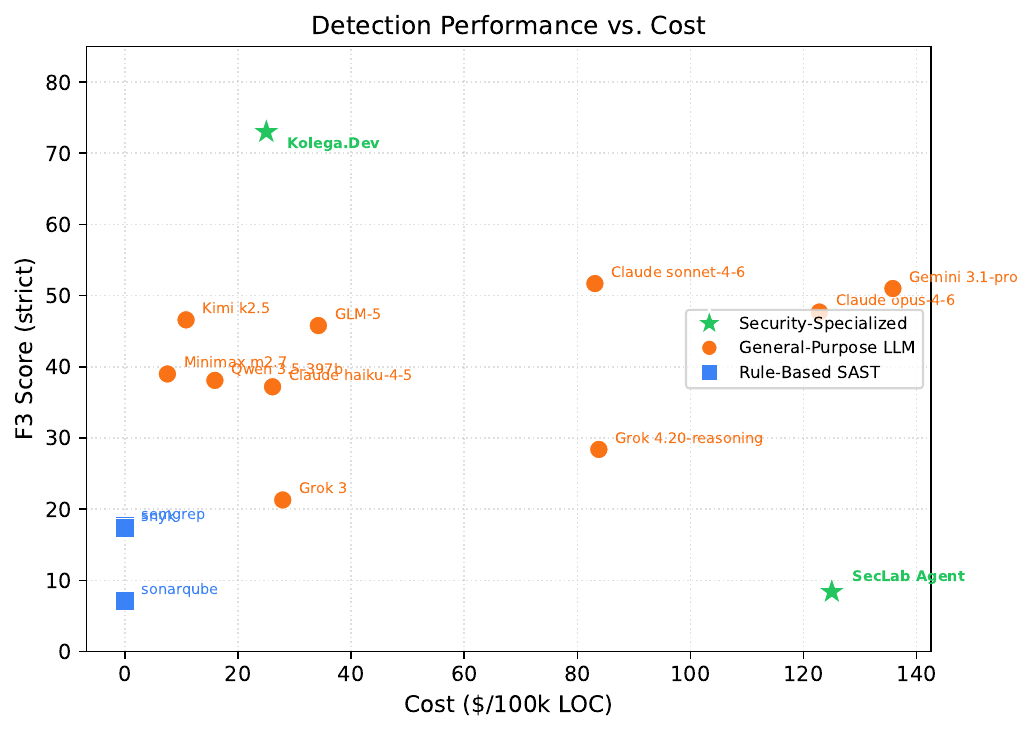}
\caption{F3 score (strict) vs.\ cost per 100k LOC. Kolega.Dev achieves the highest F3 at the lowest cost among all non-free scanners. The best GP-LLM (Sonnet) costs 3.3$\times$ more for 29\% lower F3.}
\label{fig:f3-cost}
\end{figure}

Figure~\ref{fig:f3-cost} plots detection performance against cost. Kolega.Dev achieves the highest F3 at \$25/100k LOC, occupying the upper-left region of the plot. GP-LLM scanners are scattered across the higher-cost middle, with no clear correlation between cost and performance. Gemini~3.1~Pro costs 5.4$\times$ more than Kolega.Dev for 30\% lower detection. Rule-Based SAST tools have variable pricing but detect fewer than 18\% of vulnerabilities.

\section{Analysis \& Discussion}
\label{sec:discussion}

This section interprets the comparative results across the three scanner
categories, identifies the key patterns that emerge from the evaluation, and
acknowledges the limitations and threats to validity that constrain
the conclusions we can draw.

\subsection{Key Findings}
\label{sec:key-findings}

\paragraph{A clear three-tier hierarchy emerges.}
The clearest result is the emergence of three distinct performance
tiers.  The Security-Specialized scanner (Kolega.Dev, F3\,=\,73.0 strict)
substantially outperforms even the best General-Purpose LLM (Sonnet~4.6,
F3\,=\,51.7), which in turn achieves nearly 3$\times$ the F3 score of
the best Rule-Based tool (Semgrep, F3\,=\,17.7).

This hierarchy suggests that General-Purpose LLM reasoning, while a
major advance over pattern matching, is not sufficient for
production-grade vulnerability detection.  Purpose-built security
architecture that combines LLM reasoning with domain-specific
design delivers higher recall.

Two caveats apply: the Security-Specialized tier now contains two systems, but they occupy opposite extremes: Kolega.Dev (F3\,=\,73.0) leads the benchmark while the SecLab Taskflow Agent (F3\,=\,8.4) trails it, confirming that breadth of detection coverage, not merely Security-Specialized architecture, drives top-tier performance. The GP-LLM tier exhibits high internal variance (F3 from 21.3 to 51.7), with some models performing at or below Rule-Based SAST levels.

The per-CWE breakdown reveals where these advantages originate.
General-purpose LLMs dramatically outperform Rule-Based SAST on
vulnerability classes that require semantic understanding: SQL injection
(95\% vs.\ 32\% recall), command injection (83\% vs.\ 24\%), and
insecure deserialization.  Rule-Based SAST tools remain competitive only
on issues that reduce to syntactic patterns (e.g., hardcoded secrets),
but even on those categories their overall recall remains low.

\paragraph{Security-Specialized architecture closes the gap that General-Purpose LLMs cannot.}
The 21-point F3 gap between Kolega.Dev (73.0) and the best GP-LLM (Sonnet~4.6, 51.7) reflects the advantage of domain-specific design: vulnerability-aware analysis pipelines, security-specific context assembly, and optimized detection strategies that generic prompting cannot replicate. Kolega.Dev completes 26/26 repositories with zero timeouts, whereas some frontier GP-LLMs fail to produce results for 15--27\% of repositories. The combination of higher recall \emph{and} higher reliability defines the Security-Specialized advantage.

\paragraph{Where Kolega.Dev falls short.}
Kolega.Dev's precision (0.388) is higher than both Semgrep (0.205) and Snyk (0.282), though lower than the General-Purpose LLMs (0.6--0.9) which achieve their precision by flagging far fewer findings overall. This is a deliberate design tradeoff: Kolega.Dev is optimized for F3-weighted performance, prioritizing comprehensive detection for regulated and high-risk industries (finance, healthcare, critical infrastructure) where the cost of a missed vulnerability far exceeds the cost of triaging false positives. Kolega.Dev also underperforms GP-LLMs on several individual repositories with canonical vulnerability patterns, suggesting its advantage is not universal.

\paragraph{Bigger models do not always win.}
One might expect a monotonic relationship between model capability
(and cost) and benchmark performance.  The data tell a different
story.  Opus~4.6 achieves strong per-repo scores on the repositories
it completes, but completing only 19 of 26
repos drags its strict F3 down to 47.7, below Sonnet~4.6's 51.7,
which reflects fuller coverage.  A similar pattern emerges with
Minimax, which fails to produce valid output on 31\% of runs (completing 22 of 26 repos).  Conversely,
Kimi~K2.5 and Minimax~M2.7 punch above their weight on
cost-efficiency metrics, delivering competitive F3 scores at
substantially lower per-repo cost.  The implication for practitioners is
that scanner selection should optimize for \emph{reliability-adjusted}
performance rather than peak capability: a model that reliably completes
every scan is more valuable than one that occasionally excels but
frequently fails to return results.

\paragraph{The precision--recall tradeoff is stark.}
Grok~4.20 Reasoning exemplifies the conservative end of the spectrum:
its precision of 0.927 is the highest in the benchmark, but its recall
of 0.263 means it misses nearly three-quarters of true vulnerabilities.
Most LLM scanners cluster in the 0.6--0.8 precision range with
substantially higher recall.  Among
Rule-Based tools, SonarQube illustrates the opposite failure mode:
respectable precision (0.611) paired with negligible recall (0.065),
yielding an F3 of just 7.1.  The recall-weighted F3 metric (Section~\ref{sec:scoring}) correctly rewards coverage over conservatism, aligning with
practitioner priorities where missed vulnerabilities carry far greater risk than false positives.

\subsection{Limitations}
\label{sec:limitations}

We identify five limitations that scope the conclusions of this work:

\begin{itemize}
  \item \textbf{Python only, Type~1 only.}
    Version~1.0 covers only Python and only intentionally vulnerable applications with high vulnerability density. Scanner performance may differ on statically typed or memory-unsafe languages, and on production codebases with sparser vulnerabilities. Multi-language and Type~2--5 targets are planned for v2.

  \item \textbf{Limited Security-Specialized coverage.}
    The Security-Specialized tier contains only two scanners (Kolega.Dev and the SecLab Taskflow Agent). Several other vendors market AI-powered security scanners, but we were unable to obtain access. We actively invite Security-Specialized vendors to contribute their results or reach out for integration support. The README provides a step-by-step guide for adding new scanners. Broader representation would strengthen the benchmark's conclusions about this category.

  \item \textbf{Ground truth subjectivity.}
    Hand-labeling inherently involves judgment calls. We publish labeling evidence and rationale for every entry and invite community challenges via the issue tracker.

  \item \textbf{LLM non-determinism and timeouts.}
    LLM scanners produce variable output across runs. Single-run results should be interpreted with appropriate uncertainty. Opus~4.6 completed only 19/26 repos and Minimax~M2.7 completed 22/26, reflected in the strict F3 metric.

  \item \textbf{Default configurations and missing scanner families.}
    All scanners use default settings. Tuned configurations could yield different results. Open-weight models (Llama, Mistral) are absent due to infrastructure constraints and are planned for v2. OpenAI models are represented indirectly: the SecLab Taskflow Agent uses GPT-5.4 via LiteLLM, though its multi-stage pipeline architecture means its results reflect the agent design rather than raw model capability.

\end{itemize}

\subsection{Threats to Validity}
\label{sec:threats}

\paragraph{Data contamination.}
The 26 target repositories are publicly available on GitHub, and it is
likely that some or all of them appeared in the training data of the LLM
scanners we evaluate.  If a model has memorized known vulnerabilities in
these repos, its recall could be inflated relative to performance on
unseen code.  We mitigate this threat in two ways.  First, the 120
false-positive traps in the ground truth test whether scanners
\emph{discriminate} between vulnerable and safe code. Memorizing that a
file contains vulnerabilities does not help a scanner avoid flagging a
non-vulnerable code path in the same file.  Second, the v2 roadmap
(Section~\ref{sec:living-benchmark}) includes repositories published
after model training cutoffs, which will provide a contamination-free
evaluation set.

\paragraph{Selection bias.}
Twenty-six repositories, while substantial for a hand-labeled benchmark,
do not exhaustively represent the Python vulnerability landscape.
Certain CWE families (e.g., cryptographic weaknesses, race conditions)
are underrepresented in the current corpus.  The living-benchmark
roadmap addresses this through planned expansion to additional repos
and target types, aiming for broader CWE and application-domain
coverage in future versions.

\section{The Living Benchmark}
\label{sec:living-benchmark}

A static benchmark becomes stale as soon as the scanners it measures improve.
RealVuln is designed to grow alongside the tools it evaluates.

\subsection{Versioning and Community}

Every benchmark release is identified by a deterministic \emph{manifest hash} computed from ground-truth content, pinned repository commits, and prompt versions. Scores are always reported against a named manifest version. When new repositories or corrected labels are added, the hash changes and a new version tag is created. Old results remain valid, and the live dashboard at \url{https://realvuln.kolega.dev/} surfaces side-by-side comparisons against historical versions.

RealVuln is maintained as a community resource. Contribution pathways (adding repositories, submitting scanner results, and challenging ground-truth labels) are documented in the repository README.\footnote{\url{https://github.com/kolega-ai/Real-Vuln-Benchmark}} The live leaderboard is rebuilt automatically on every merge to the \texttt{deploy\_pages} branch.

\subsection{Roadmap}

The following extensions are planned for v2:
\begin{itemize}
  \item \textbf{Multi-language expansion:} JavaScript/TypeScript, Go, and Java repositories.
  \item \textbf{Type~2 targets:} production apps with CVE-linked commits at realistic vulnerability density.
  \item \textbf{Type~3 targets:} vulnerable open-source libraries patched for disclosed CVEs.
  \item \textbf{Type~4 roll-ups:} OWASP Benchmark and Juliet re-scored under RealVuln metrics.
  \item \textbf{Agentic sandbox:} fully isolated, reproducible agentic evaluation via Docker containers.
\end{itemize}

\section{Conclusion}

We presented a systematic comparison of Rule-Based SAST tools, General-Purpose
LLM scanners, and Security-Specialized systems, evaluated on
\textsc{RealVuln}, the first fully open-source benchmark built from real-world
vulnerable code.  Version~1 spans 26 repositories, 796 manually labeled
findings, and 120 false-positive traps across 15 scanners in three categories.
Every artifact (ground truth, raw scanner output, scoring pipeline, dashboard
generator, and evaluation prompts) is released under a permissive license
with a reproducible manifest that version-locks results to exact benchmark
snapshots.

Our experiments reveal a clear three-tier hierarchy.  The Security-Specialized
scanner (Kolega.Dev, F3\,=\,73.0 strict) leads the benchmark, demonstrating that
purpose-built security architecture delivers meaningfully higher recall than
General-Purpose approaches.  General-purpose LLM scanners form a competitive
middle tier. The best, Claude Sonnet~4.6 (F3\,=\,51.7), achieves nearly
3$\times$ the F3 score of the best Rule-Based SAST tool (Semgrep, F3\,=\,17.7).
Nevertheless, no scanner dominates across all CWE families: Rule-Based tools
retain a precision advantage on well-specified injection patterns, while
LLM-based systems excel on logic-level and authentication flaws that require
broader code context.  The contrast between the two Security-Specialized systems reinforces this conclusion: the SecLab Taskflow Agent, despite sophisticated multi-stage threat modeling with GPT-5.4, scores F3\,=\,8.4, below every Rule-Based SAST tool, because its depth-first design audits only 3--5 threat categories per repository. Purpose-built breadth of coverage, not merely Security-Specialized architecture, is what separates top-tier detection from the rest.
These results suggest that the future of vulnerability
detection lies not in applying General-Purpose AI to security prompts, but in
building Security-Specialized systems that combine LLM reasoning with
domain-specific architecture \emph{and} broad detection coverage.

The benchmark is designed to evolve.  A community contribution pathway,
described in \S\ref{sec:living-benchmark}, allows new repositories, scanners,
and ground-truth corrections to be incorporated without invalidating prior
results.  Version 2 will expand coverage to JavaScript/TypeScript, Go, and
Java, and will introduce Type 2 targets drawn from real production CVEs,
setting a higher bar for precision in realistic, low-vulnerability-density
codebases.  We invite the security research community to submit scanner results,
propose new repositories, and challenge existing labels.

All code, data, results, and tooling are available at \url{https://github.com/kolega-ai/Real-Vuln-Benchmark}. A live interactive dashboard is hosted at \url{https://realvuln.kolega.dev}. We encourage the security community to verify our results, challenge our ground truth, and submit their own scanner evaluations.

\bibliographystyle{plainnat}
\bibliography{references}

\end{document}